\documentclass[12pt]{iopart}
\usepackage{graphicx}

\begin{document}

\title[]{Pseudo-Phase Transitions of Ising and Baxter-Wu Models in Two-Dimensional Finite-Size Lattices}

\author{Wei Liu$^1$, Fangfang Wang$^1$, Pengwei Sun$^1$, and Jincheng Wang$^1$}

\address{$^1$College of Science, Xi'an University of Science and Technology, Xi'an, China}
\ead{weiliu@xust.edu.cn}
\vspace{10pt}
\begin{indented}
\item[] 
\end{indented}

\begin{abstract}
This article offers a detailed analysis of
pseudo-phase transitions of Ising and Baxter-Wu models in two-dimensional
finite-size lattices. We carry out Wang-Landau sampling to obtain the density of states.
Using microcanonical inflection-point analysis with microcanonical entropy, we
obtain the order of the psuedo-phase transitions in the models. The microcanonical analysis results of the
second-order transition for the Ising model and the first-order transition for the Baxter-Wu model are
consistent with the traditional canonical results. In addition, the third-order transitions are
found in both models, implying the universality of higher-order phase transitions.
\end{abstract}

%
\vspace{2pc}
\noindent{\it Keywords}: pseudo-phase transitions, Baxter-Wu model, microcanonical inflection-point analysis
%
%
%
%

\section{Introduction}

Traditionally, phase transitions refer to analytical discontinuities or singularities in the thermodynamic
functions of systems that satisfy the thermodynamic limit \cite{Pathria}. Landau's phenomenological
theory successfully describes all second-order phase transitions in a mean-field sense. In 1971, Wilson
introduced the concept of a renormalization group into the theory and interpreted the scaling laws and
the universality class \cite{wilson1983renormalization}. In the construction of the theory, the Ising model, which could be precisely solved, 
played a crucial role \cite{brush1967history}. This model provides the criteria for judging the precision of theories for phase transitions.

Generally, models that are more complicated than the Ising model could not be precisely solved. As powerful tools, Monte Carlo
simulations offer numerical methods for calculating thermodynamic functions \cite{landau2021guide}.
Although computer simulations cannot be conducted on infinite systems, one can obtain the
precise critical behaviors and transition temperatures via finite-size analysis.
By changing the size of the system, one can obtain the scaling behavior of the singular part of the free
energy as well as the thermodynamic properties. The majority of the existing simulation studies have revealed the behaviors
under the thermodynamic limit. 
 However, with the development of nano-science, biophysics and network
science, many small systems have drawn a tremendous amount of attention from researchers \cite{chamberlin2015big,chamberlin2003critical,newman2018networks}.
 We notice that these finite systems still show some dramatic changes in their "macrostate properties".
There is no singularity in such systems; we cannot use the theories for large systems to 
deal with small systems. Therefore, how to understand the behaviors
of phase transitions in finite systems is an important question. 

Researchers have studied the psuedo-transitions in small systems using microcanonical analysis \cite{gross2001microcanonical}.
Using the Microcanonical  Metropolis Monte Carlo method, one can obtain the first-order phase transitions
in the Potts model and the second-order transitions in the Ising model.
Recently, Qi and Bachmann \cite{qi2018classification} generalized the microcanonical inflection-point analysis method to identify
and locate independent and dependent phase transitions of any order for finite-size systems. 
This method has been applied to the study of flexible polymers, and has been proven to be very successful \cite{qi2019influence}.
The results of these studies reveal that microcanonical inflection-point analysis can help 
identify the major transitions and distinguish the details of the transition
processes by signaling higher-order transitions. Moreover, these researchers found the dependent transitions which
can only occur in coexistence with independent transitions of a lower order.
Using the exact density of the state (DOS) of the Ising model \cite{beale1996exact}, Sitarachu et al. studied the 1D and 2D finite-size
Ising model in detail \cite{sitarachu2020exact}. The high-order phase transitions were confirmed in the 2D Ising system.

D. W. Woods and H. P. Griffiths \cite{wood1972self} proposed a magnetic model (the Baxter–Wu (BW) model), 
which is defined in a two-dimensional triangular lattice. R. J. Baxter and F. Y. Wu \cite{baxter1973exact,baxter1974ising}
precisely solved the spin-1/2 model and found that it belongs to the four-state Potts universality class. 
Moreover, the transitions of the finite-size systems of the
BW model displayed discontinuities due to low frequencies and large energy fluctuations \cite{schreiber2005monte,liu2021phase}.
In addition to being applied to the fields of magnetism and surface adsorption, the BW model
was used to study the dynamics of social balance \cite{antal2005dynamics} and the satisfiability problem of computer
science \cite{radicchi2007social} in the field of complex networks.
The order of the dynamical and thermodynamic phase transitions and the transition points were obtained
when the size of the social systems went infinity\cite{kargaran2021heider,masoumi2021mean,rabbani2019mean,krawczyk2021structural}.
However, social systems are usually finite. It is necessary to study the pseudo-phase transitions for finite systems in
spin models to reveal the universal behaviors of the real systems.

The canonical methods \cite{schreiber2005monte} pointed out that the
internal energy histograms of the single-layer 2D spin-1/2 BW model 
have double peaks, which suggest
pseudo-first-order phase transitions at the finite-size systems. The
reason for this phenomenon is that the system is in a metastable state of coexistence
with the ferro- and ferri-magnetic orders. 
Moreover, Jorge et al. \cite{jorge2020order} investigated the spin-1 phase transition of the Baxter-Wu model and found more
complicated behaviors of which the system undergoes a tetracritical transition, with the coexistence of a 
ferromagnetic and three ferrimagnetic states. By observing the canonical distributions of the
internal energy near the transition points, canonical methods can give the information of pseudo-first-order
phase transitions and continuous phase transitions. However, the methods could not systematic classify the
pseudo-phase transitions, and they could not locate the pseudo-transition points precisely due to the different
maximums of the different thermodynamic quantities. Fortunately, generalized microcanonical inflection-point analysis
can do these works \cite{qi2018classification}.

It is necessary to study the finite BW model using the perspective of microcanonical inflection-point analysis.
In this article, we use the Wang-Landau method to obtain the DOSs of the Ising model and the Baxter--Wu model.
The Ising model is used to certify the correctness of the DOS obtained via Wang--Landau sampling. In Section 2, we
briefly describe the models and their methods. The results are shown in Section 3, along with some discussion.
Section 4 provides the conclusions of our work.

\section{The model, simulation and data analysis methods}

\subsection{Ising model and the Baxter-Wu model}

The Hamiltonian of the Ising model built on the $N \times N$ square lattices is as follows,
\begin{equation}\label{hamiltonianising}
E = -J\sum _{ <ij>}s_{i}s_{j} 
\end{equation}
where $s_{i}=\pm 1$ stands for the spin located in the lattices, and $<i,j>$
denotes the summations over the nearest neighbor sites. We only investigated
the ferro-magnetic model, thus $J>0$. If one spin of the lattice flips, it leads to
the $4J$ energy difference in most cases. There are two exceptions in which the differences
between the ground state and the first excited state, and the highest energy level and
the second-to-last energy level are $8J$. As a result, there are $N^{2}-1$ energy
levels in this model.

The Hamiltonian of the BW model in the $N \times N$ triangular lattices is
\begin{equation}\label{hamiltonianbw}
E = -J\sum _{<ijk>}s_{i}s_{j}s_{k}
\end{equation}
where the spin variables are located at the vertices of the lattice.
$<ijk>$ denotes the product of the three spins on the elementary triangle, and the sum is over all triads $<ijk>$.
We also studied the case of $J>0$. For spin-$1/2$, one spin flip also leads to a $4J$ energy difference in most cases.
However, the differences between the ground state and the first excited state, and highest energy level and
second-to-last energy level are $12J$. Therefore, the spin-1/2  model has $N^{2}-3$ energy
levels. If we consider the spin-$1$ BW model, in which the spin can take $-1$, $0$, or $1$,
the energy levels are more complicated; there are
$4N^{2}-10$ levels. The reason for this behavior is that the energy differences of the first and the last seven levels
are $6J$, $4J$ or $2J$, whereas the other differences are $J$.

\subsection{Wang--Landau sampling}

We carried out Wang-Landau sampling \cite{wang2001efficient} to obtain the density of states $g(E)$.
The Wang-Landau sampling method is a powerful algorithm used for the direct estimation of $g(E)$ by
random walking in the energy space with a flat histogram. 
We started with $g(E)=1$, and improved it in the following way. The random walk
was performed according to the probability, as follows:
\begin{equation}
p(E_{1}\rightarrow E_{2})=\min(\frac{g(E_{1})}{g(E_{2})}, 1) \label{wltr}
\end{equation}
where $E_{1}$ and $E_{2}$ are the energy levels before flipping, which would be the results
if the spin were flipped. The density of states was updated by
\begin{equation}
g(E)\rightarrow g(E)f
\end{equation}
where $E$ is the energy level of the accepted state, and $f$ is a modification factor,
the initial value of which is $f=f_{0}=e=2.71828$.
Meanwhile, the energy histogram is $H(E)$ plus $1$. We proceeded with the random
walks until the histogram of the energies was "flat", which, in this paper, was when the histogram for
all possible energies was no less than $80\%$ of the average histogram. Then, the modification factor was reduced,
$f_{i+1}=f_{i}^\frac{1}{2}$, and we reset the histogram to $H(E)=0$ for all values of $E$, and
restarted the random walks. The simulation was stopped when the modification factor
was smaller than $f_{final}=1+10^{-8}$. 

The advantage of this method is avoiding the critical slow down.
However, sampling the full range of energies of a large system with a complex energy
landscape is quite difficult using a single walk \cite{cunha2008improving}, since
obtaining a  "flat" histogram of the energies requires a very long simulation time.
Vogel et. al. suggest the replica-exchange Wang–Landau sampling method 
\cite{vogel2013generic,vogel2014scalable,vogel2014exploring,vogel2018practical},
which is suitable for studying a large system with  a long energy range and complex landscape.
This algorithm splits the energy range into many smaller sub-windows with large overlap between adjacent windows.
We carried out WL sampling in each energy sub-window with multiple independent walkers.
A $75\%$ overlap was chosen in this work.
After a certain number of Monte Carlo steps, a replica exchange was proposed between two random walkers.
Suppose $i$ and $j$ are random walkers of two adjacent energy windows, and $\{s_{i}\}$ and $\{s_{j}\}$ are the configurations that the 
walkers $i$ and $j$ are carrying before the exchange. The acceptance probability $P_{acc}$ for the exchange of configurations 
 $\{s_{i}\}$ and $\{s_{j}\}$  between walkers $i$ and $j$ is
 \begin{equation}
 P_{acc} = \min \left[ 1, \frac{g_{i}(E(\{s_{i}\})g_{j}(E(\{s_{j}\})}{g_{i}(E(\{s_{j}\})g_{j}(E(\{s_{i}\})} \right]
 \end{equation}
 In order to calculate a single g(E) over the entire energy range, we chose the joining point for any two overlapping densities of states'
 pieces where the inverse microcanonical temperatures $\beta = d \log[g(E)]/dE$ best coincide.
The convergence of $g(E)$ was sped up for the smaller energy range of the sub-window.

\subsection{Microcanonical inflection-point analysis method}

Qi and Bachmann combined the microcanonical inflection-point analysis method and
the principle of minimal sensitivity to identify and classify first- and higher-order
transitions in complex systems of any size \cite{qi2018classification}.
Using this analysis, they found that the two-dimensional ferro-magnetic Ising
model exhibits signals of transitions other than the single second-order phase transition.
Third-order independent and dependent transitions were observed in the systems.
They also obtained the phase diagram for a grafted lattice
polymer interacting with an adhesive surface.

The basic idea is that the macroscopic behaviors of a physical system are
governed by the quantities of entropy and energy. The microcanonical entropy,
which contains the complete information about the phase behavior of a system,
can be defined as
\begin{equation}
S(E)=k_{B}\ln{g}(E)
\end{equation}
where $k_{B}$ is the Boltzmann constant. The entropy and its derivatives are
monotonic functions within energy regions associated with a single phase. 
However, a phase transition would break the monotony, and is singled by an inflection
point. According to the principle of minimal sensitivity, only least-sensitive inflection
points have a physical meaning. For example, if $S(E)$ has a least-sensitive inflection
point, its first derivative, $\beta(E)=T^{-1}(E)=\frac{dS(E)}{dE}$, which is the
microcanonical inverse temperature, should contain a positively valued maximum slope.
$\gamma(E)=\frac{d\beta(E)}{dE}=\frac{d^{2}S(E)}{dE^2}$ has a positive maximum,
which gives a signal of the first-order phase transitions.  If $\beta(E)$ has a least-sensitive inflection
point, its first derivative, $\gamma(E)$,
should contain a negative maximum, which reveals a second-order phase transition.
$\delta (E)=\frac{d\gamma(E)}{dE}=\frac{d^{3}S(E)}{dE^3}$ would provide the information
of a third-order phase transition.

In addition to the class of independent transitions, there is another category, dependent
transitions, which can be identified by this method as well. We summarized this method in Table \ref{method}.
\begin{table}
\centering
 \caption{Signal of the order of the transitions.\label{method}}

\begin{tabular}{@{}lll}
\br
 Categories & Even order transitions & Odd order transitions \\  
 \mr
 Independent & $\frac{d^{2k}S(E)}{dE^{2k}}<0$ & $\frac{d^{2k-1}S(E)}{dE^{2k-1}}>0$ \\    
 & Negative maximum & Positive minimum \\  \hline
  
Dependent& $\frac{d^{2k}S(E)}{dE^{2k}}>0$ & $\frac{d^{2k-1}S(E)}{dE^{2k-1}}<0$ \\  
& Positive minimum & Negative maximum \\ 
\br
 \end{tabular}
 \end{table}

Direct derivatives 
with discrete microcanonical entropy will be affected by the noise
associated with the numerical error of the data. To avoid the noise, we used the two-step
strategy. The first step is straightforward. We computed the DOS ten times and then
averaged the DOS to obtain a smoother curve. Next, we used the B\'ezier algorithm to generate
the smooth function \cite{janke2002statistical}.
The discrete data points $f_{q}$ at energy $E_{q}$ were obtained in the Wang-Landau
simulations and the derivatives of the microcanonical entropy were the control point 

\begin{equation}
f_{bez}=\sum^{Q}_{q=0}C^{q}_{Q}\left(\frac{E_Q-E}{E_Q-E_0}\right)^{Q-q}
             \left(\frac{E-E_0}{E_Q-E_0}\right)^{q}f_{q}
\end{equation}
where $q$ and $Q$ are the energy level and the highest level, respectively, and $E_0$
is the energy of the ground state. $f_{q}$ denotes $S_{q}$, $\beta_{q}$, $\gamma_{q}$
or $\delta_{q}$.
The derivatives required for the statistical analysis were calculated from this function
in this paper. We also used the independent computations ten times 
 to
obtain the error bars. If error bars were not seen, they were smaller than the size of the symbols.

\section{Results}
The purpose of this work is to identify the order of the phase transitions
for the finite Ising and BW systems. To certify the correctness of our work,
we calculated the value of the specific heat and compared it with $\gamma(E)$.
Then, the curves of $\delta(E)$ were used to locate the third-order transitions for the
models studied in our work.  In order to compare the curves of the 
different lattice sizes, we use energy per site $e=E/N^2$ to draw the figures. 

\subsection{Ising model}
\begin{figure}
\includegraphics[width=0.9\textwidth]{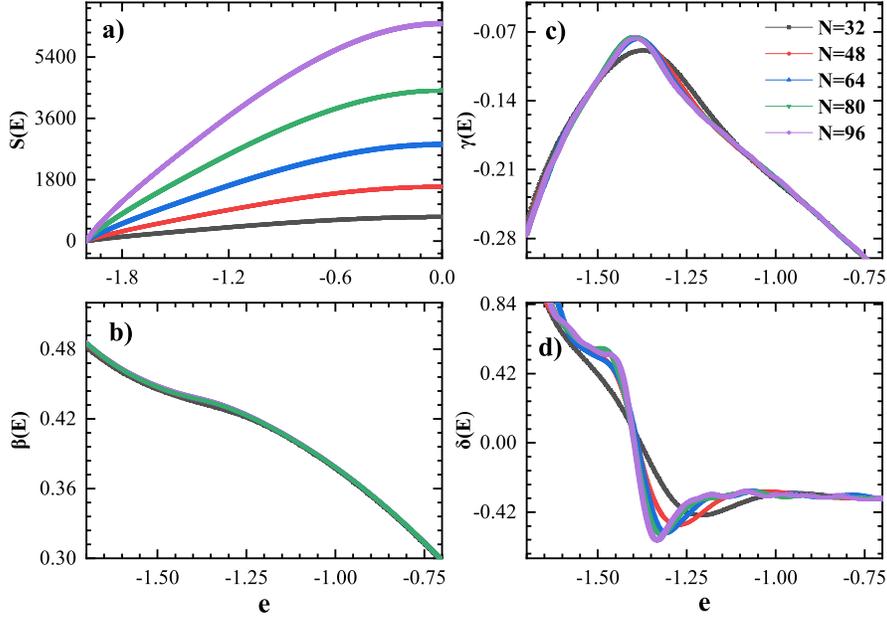}
 \caption{ a) The logarithm of the density of states; b) the inverse temperature, $\beta(e)$;
 the first and the second derivatives of the inverse temperature, c) $\gamma(e)$ and d) $\delta(e)$
 for the Ising model. \label{Ising_micro}}
 \end{figure}

Microcanonical inflection-point analysis method uses derivatives which are sensitive to
small statistical fluctuations of DOS. Though the replica-exchange
Wang–Landau sampling method can simulate the lattice size up to $256 \times 256$ \cite{vogel2013generic},
there usually exist relative larger simulation noise for large lattice sizes.
The lattice sizes we simulated were from $N=32$ to $N=96$. The microcanonical entropy
and the derivatives are shown in Fig. \ref{Ising_micro}. There are negative
maximums of the curves of $\gamma (E)$, which suggest second-order phase transitions.
We also located the transition points using the canonical method. With the DOS 
in hand, we could calculate the specific heat via  $c_{V} = \frac{<E^2>-<E>^2}{N^{2}k_{B}T^2}$,
where $<E^n> =\sum_{E}E^n g(E)\exp(-E/k_{B}T)$. The transition points are located
with the maximums of the specific heat.
The second row provides the points obtained by the microcanonical analysis and the third
row provides the results obtained by the specific heat. The results are in good agreement with each other. The results are displayed in Table \ref{locations}. 
\begin{table}
\centering
 \caption{Locations of the second-order transitions for the Ising model obtained by the microcaonical and canonical methods.\label{locations}}
\begin{tabular}{@{}llll}
\br
 $N$ & 32 & 64 & 96  \\
\mr
 $1/\beta_{micro}$ & 2.308 & 2.287 & 2.281 \\ 
  
 $1/\beta_{canonical}$ & 2.296 & 2.281 & 2.277 \\ 
 \br
\end{tabular}
 \end{table}

\begin{figure}
\includegraphics[width=0.9\textwidth]{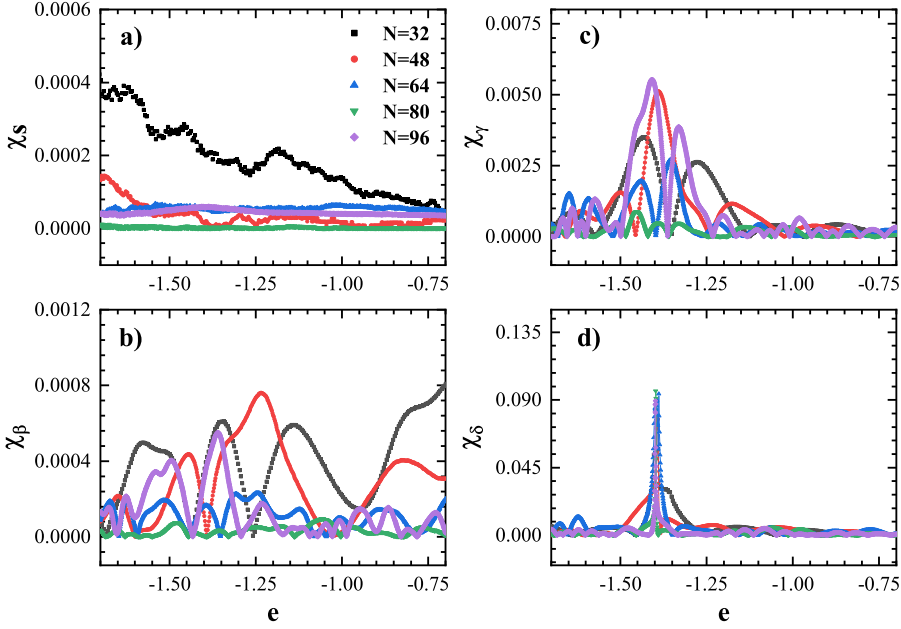}
\caption{ Relative errors of a) the logarithm of the DOSs; b) the inverse temperature, $\beta(e)$;
the first and the second derivatives of the inverse temperature, c) $\gamma(e)$ and d) $\delta(e)$
for the Ising model.\label{Ising_compare}}
\end{figure}

In order to further verify the correctness of the DOS obtained by Wang-Landau sampling for
the microcanonical analysis, the results from the exact DOS  \cite{Pathria,beale1996exact}
and its derivatives are compared with Wang-Landau data. The relative errors are displayed
in Fig. \ref{Ising_compare}. The systematic
error of the microcanonical entropy is very extremely small, up to $10^{-5}$ to
$10^{-4}$ and the random error is smaller than the systematic one.
The higher order derivatives certainly enlarge the noises as showed in Fig. \ref{Ising_compare} c) and d).
Fortunately, the results of the third order derivative of microcanonical entropy $\delta (E)$ could be considered reasonable.

$\delta (E)$ shows the positive minimum in Fig.\ref{Ising_micro} when $N = 96$, which suggests the independent third-order
transitions. The location is $e = -1.469$. Moreover, when the size was small, there were
infection-points suggesting fourth-order transitions. However, we could not locate the points precisely
because the higher order derivatives amplified the noise of the simulation.

The third dependent transition signals were detected as well. The curves for
$\delta(E)$ had the negative maximums which could determine the third transition points.
The points are shown in Table \ref{third}.
\begin{table}
\centering
\caption{Locations of the dependent third-order phase transitions for the Ising model. \label{third}}
\begin{tabular}{@{}lllll}
\br
$N$ & 32 & 48 & 64 & 96  \\  
\mr

 $1/\beta$ & 2.727 & 2.597& 2.554 & 2.540 \\  
\br
\end{tabular}
\end{table}
\unskip

The transition points obtained with the microcanonical and canonical methods coincide well and are in agreement
with the results of Qi et al.'s work \cite{qi2018classification}, which shows that it is reliable
to perform microcanonical inflection-point analysis up to third-order transitions using the DOS obtained by the
Wang--Landau sampling method. 

\subsection{Baxter-Wu model}
The BW model in the triangular lattice has four degenerate ground states,
and has a longer energy range for the spin-1 model. Hence, the simulated lattice
sizes are from $N=30$ to $N=90$ for the spin-1/2 model and from
$N =18$ to $N=60$ for the spin-1 model.

\begin{figure}
\includegraphics[width=0.9\textwidth]{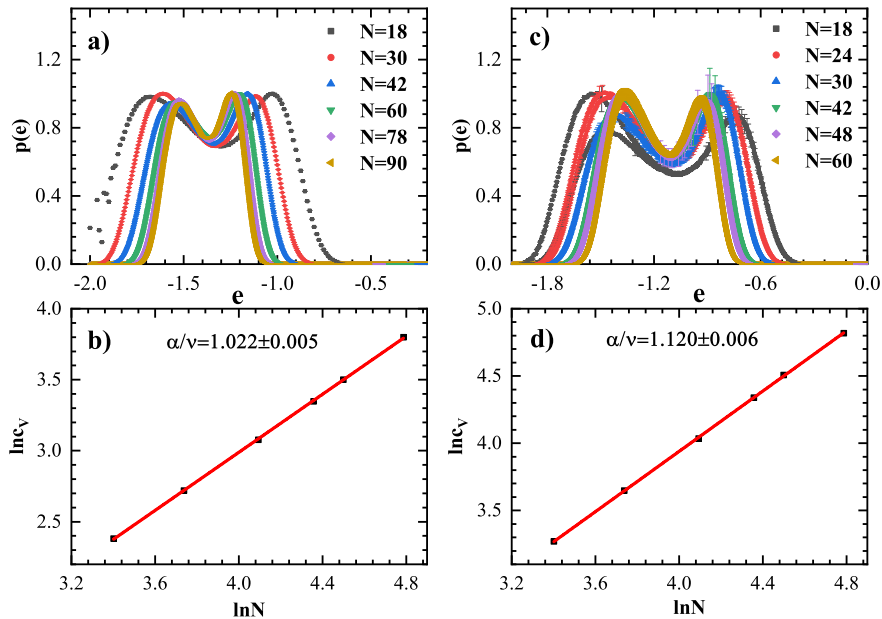}
 \caption{The probability distributions of
 the internal energy and the finite-size effect of the specific heat
 of the Baxter--Wu model: a), b) for spin-1/2, and c), d) for spin-1.  \label{bwpe}}
 \end{figure}

\begin{figure}
 \centering
\includegraphics[width=0.8\textwidth]{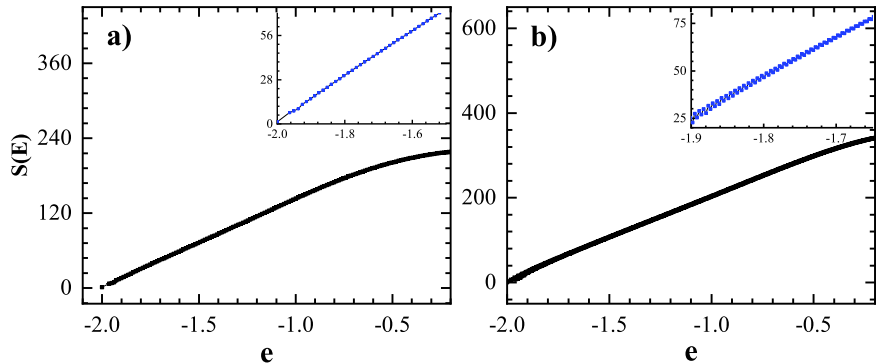}
 \caption{The DOSs of the BW model at lattice size $N=18$, a) for spin$-1/2$, b) for spin-1.
 The insets are the enlarged views of the low energy regions.\label{dos18}}
 \end{figure}

Using our obtained data on the DOS, we calculated the specific heat and analyzed the finite-size effect.
The result we obtained is that the critical exponent $\alpha/\nu$ is equal to $1.022\pm 0.005$ for the spin-1/2 model, whereas it is equal
to $1.120\pm 0.006$ for the spin-1 model.
The spin-1/2 model belongs to the universality class of four-state Potts, of which the exponent nearly equals
$1$. However, the exponent for the spin-1 model is larger than $1$, which suggests that
the behavior of the thermodynamic limit of this model is nontrivial. Fig. \ref{bwpe} gives
the finite-size effect and the energy distribution of the BW model.
The double-peak distribution suggests the first-order transition. 
Besides, the distribution of spin-1/2 points at low energies shows "random" behavior to some
extent when lattice size is small. We emphasize here that the behavior does not come from simulation noise,
and the behavior is not random.
The DOS at low energy of this model results in the effect.
Fig. \ref{dos18}. a) displays the DOS at $N=18$, and the error bars are too small to see.
The DOS for spin-1/2 model shows the distinct non-smoothness, which may lead to the deterministic
"random" distribution. 

In contrast to the case of the spin-1/2 model,
the distribution at a small lattice size for the spin-1 model shows two different height peaks at a low energy region.
The DOS of $N=18$ (Fig. \ref{dos18} . b)) for spin-1 model shows fluctuations that the number of the
states at lower energy is larger than the states at nearest higher energy levels at the region of
the low energy level, which gives rise to the two peaks.
The reason for the unexpected phenomena is that square lattice adding next-nearest-neighbour diagonal bonds
is used to produce a triangular lattice. The shape of this representation of triangular lattice
is of a rhombus. As a result, for smaller lattice size, it can give rise to unexpected finite-size
effect due to the different rotational symmetries from the real triangular lattice \cite{newman1999monte}.

\begin{figure}
\includegraphics[width=0.9\textwidth]{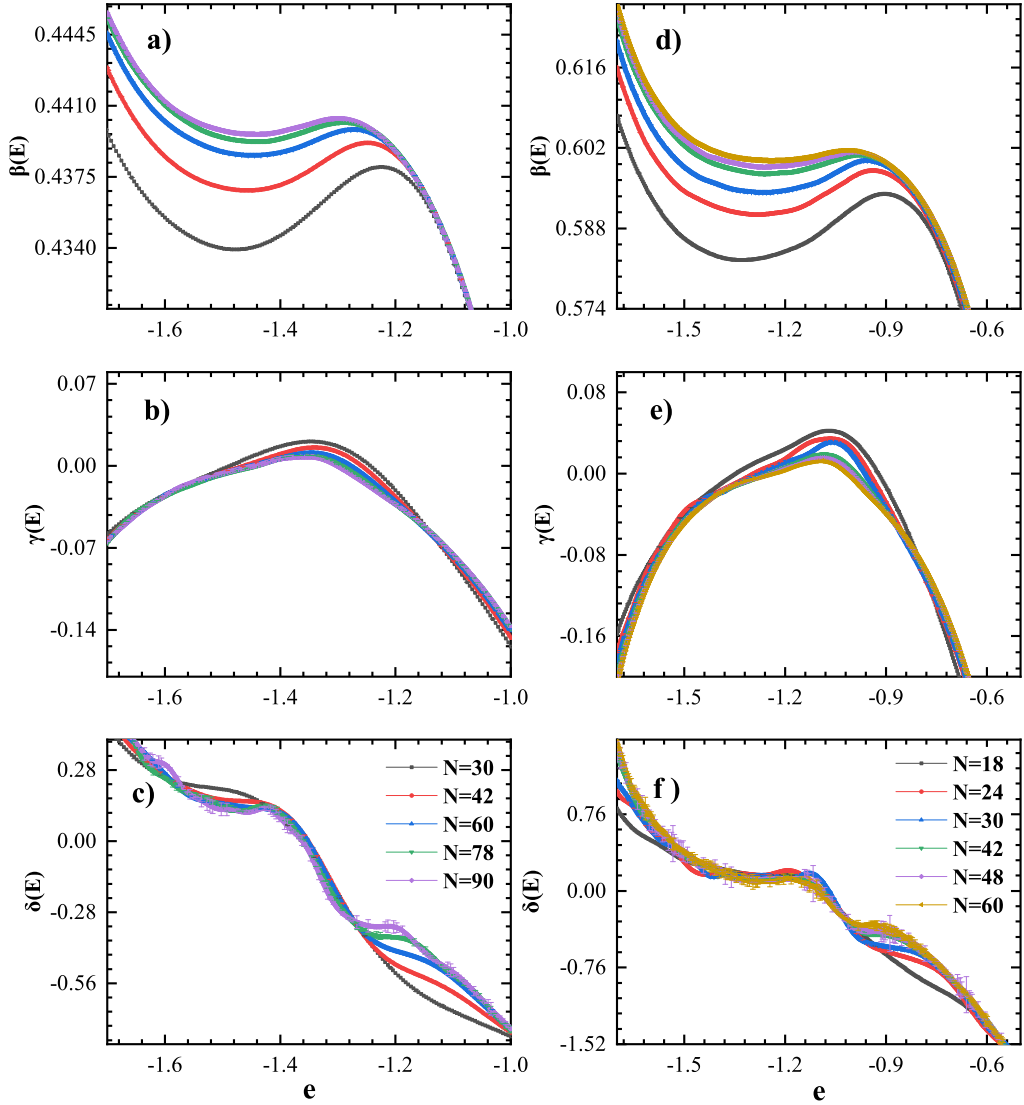}
 \caption{The inverse temperature $\beta(e)$,
 the first and the second derivatives of the inverse temperature $\gamma(e)$ and $\delta(e)$; 
 a)--c) for spin-1/2, d)--f) for the spin-1 BW model. \label{bwsub}}
 \end{figure}

The microcanonical results for the spin-1/2 and  spin-1 BW models are displayed in Fig. \ref{bwsub}.
The non-monotonic "backbending" of the inverse temperature
$\beta$, which is a typical signal of phase coexistence, is observed in
both the spin-1/2 and spin-1 models. The positive maximum values of $\gamma$
reveal the first-order phase transitions of which the positions locate the
transition points. As the lattice size increases, the values of $\gamma$ become smaller,
implying the decrease in the "latent heat". This result is consistent with the results of
the canonical analysis shown in Fig \ref{bwpe} a) and c).

Moreover, the signals of the third-order phase transitions
appear in the curves of Fig. \ref{bwsub} c and f. Similar to the Ising model, both the independent and
dependent transitions were detected. Some of the curves of $\delta(E)$ had positive
minimums and negative maximums. However, if the lattice sizes are small,
positive minimums and negative maximums of the curves may not appear.
The behaviors are found in Ising model as well. The smaller system might show
fourth-order phase transitions  \cite{qi2018classification} which
are hardly detected via Wang-Landau sampling data due to the simulation noises.
Table \ref{thirdbw} gives the locations of the third-order
phase transitions. $\beta_{a,b}$ stands for the transition inverse temperature, where
$a$ denotes the spin-1/2 or 1 models 
 and $b$ denotes the type of transition. $d$ is for the dependent transition
and $in$ is for the independent transition. NS stands for "not simulated" and
NF for "not found".

\begin{table}
\centering
\caption{Locations of the third-order transitions for the BW models. \label{thirdbw}}

\begin{tabular}{@{}lllllllll}
\br
  $N$ & 18 & 24 & 30 & 42 & 48 & 60 & 78 & 90  \\  
 \mr

  $1/\beta_{1/2,d}$& NS & NS & NF& NF & NF & NF & NF & 2.277 \\  
  $1/\beta_{1/2,in}$&  NS & NS& NF & NF & NF& NF & 2.276 & 2.274\\ 
  $1/\beta_{1,d}$& NF & NF & NF & NF & 1.665 & 1.664 & NS & NS \\   
  $1/\beta_{1,in}$&  1.715 & 1.688& 1.682 & 1.673 & 1.670 & 1.667 & NS & NS\\  
 \br 
\end{tabular}
\end{table}

Due to the existence of the latent heat of the pseudo-first-order transitions, the higher-order
phase transition shows some differences from the system with the second-order transition.
The positions of the third-order independent phase transitions are at lower energy levels than the positions of
the first-order phase transitions, whereas the positions of the dependent phase transitions are at higher energy levels.
The inverse temperature $\beta$ has a non-monotonic "backbending" phenomenon, and the locations
of the third transitions are in the region of the backbending, which leads to the
close positions between the independent and dependent transitions. In addition, the independent
transitions refer to the transitions between the over-heated states with higher orders and the metastable states
with low orders, whereas the dependent transitions might refer to the transitions between supercooled lower-order
metastable states and the disordered states. The third-order phase transitions occur in the region of the
coexisting phase, hence, the positions of the independent and dependent transitions
are very close.

\section{Conclusion}

In summary, we used Wang--Landau sampling to obtain the DOSs for the Ising
and BW models. Then, the orders of the pseudo-transitions for the finite-size systems were
analyzed via the microcanonical inflection-point method. We also compared the microcanonical
results with the canonical ones. The data for the Ising model reveal that
the DOS obtained via Wang--Landau sampling can be used for analyzing the order of transitions up
to the third order, and the positions of the second- and third-order transitions
were precisely obtained. The differences of the phases below or above critical point were
detected with the independent or dependent third-order pseudo-phase transitions which
could not be observed via canonical methods.
Based on the reliability of this method, we studied the spin-$1/2$
and -$1$ BW models. The data show that the finite-size BW models have the first and
third transitions. The "latent heat" of the first-order transitions decreases as the
systems become larger in size. We conjecture that the "latent heat" disappears at the
thermodynamic limit. Third-order phase transitions in both the
Ising and BW models imply the universality for the higher-order transitions to some extent.
The dependent third order transitions occur at a higher energy than the corresponding independent transitions,
which reveals that the systems would be in a less ordered phase before it goes into a
ordered phase. This might have potential applications in materials science since dependent
higher order transitions might help ones find the unstable structures.
Consequently, it would be of importance to confirm the universality of the higher order phase transitions in
different spin models and reveal the nature of transitions higher than the second order in future work.

\section{Acknowledgements}
We would like to thank Prof. Michael Bachmann for his valuable discussion
and thank the anonymous referee for his/her helpful suggestions.
This work is supported by the Natural Science Basic Research Program of Shaanxi
(Program No.2022JM-039)
\vspace{10pt}

\bibliographystyle{iopart-num}
\bibliography{sample}

\end{document}